\documentclass[conference]{IEEEtran}
\IEEEoverridecommandlockouts
\usepackage{cite}
\usepackage{amsmath,amssymb,amsfonts}
\usepackage{algorithmic}
\usepackage{graphicx}
\usepackage{textcomp}
\usepackage{xcolor}
\usepackage{calrsfs}
\usepackage{balance}
\usepackage{caption}
\DeclareMathOperator*{\argmax}{arg\,max}

\def\BibTeX{{\rm B\kern-.05em{\sc i\kern-.025em b}\kern-.08em
    T\kern-.1667em\lower.7ex\hbox{E}\kern-.125emX}}

\begin{document}

\title{Rethinking Maximum Flow Problem and Beamforming Design through Brain-inspired Geometric Lens
}
\author{\IEEEauthorblockN{Ahmed S. Ibrahim}
\IEEEauthorblockA{Department of Electrical and Computer Engineering, 
Florida International University,
Miami, Florida, USA \\
aibrahim@fiu.edu}
}

\maketitle
\vspace{-1.4in}
\begin{abstract}
Increasing data rate in wireless networks can be accomplished through a two-pronged approach, which are 1) increasing the \emph{network flow rate} through parallel independent routes and 2) increasing the \emph{user's link rate} through beamforming codebook adaptation. Mobile relays are utilized to enable achieving these goals given their flexible positioning. First at the network level, we model regularized Laplacian matrices, which are symmetric positive definite (SPD) ones representing relay-dependent network graphs, as points over Riemannian manifolds. Inspired by the geometric classification of different tasks in the brain network, Riemannian metrics, such as Log-Euclidean metric (LEM), are utilized to choose relay positions that result in maximum LEM. Simulation results show that the proposed LEM-based relay positioning algorithm enables parallel routes and achieves maximum network flow rate, as opposed to other metrics (e.g., algebraic connectivity). 

Second at the link level, we design unique relay-dependent beamforming codebooks aimed to increase data rate over the spatially-correlated fading channels between a given relay and its neighboring users. To do so, we propose a \emph{geometric machine learning} approach, which utilizes support vector machine (SVM) model to learn an SPD variant of the user's channel over Riemannian manifolds. Consequently, LEM-based Riemannian metric is utilized for classification of different channels, and a matched beamforming codebook is constructed accordingly. Simulation results show that the proposed geometric-based learning model achieves the maximum link rate after a short training period.  
\end{abstract}

\vspace{0.05in}

\vspace{-0.07in}
\section{Introduction} \label{sec_intro}
Mobile networks aim to support the emerging \emph{high-rate} applications (e.g., virtual/augmented reality) and also be \emph{adaptive} to spatio-temporal variability in wireless traffic demands~\cite{Kibilda_spatio-temporal_traffic}. \emph{Dynamic} positioning of mobile relays, such as vehicular road side units (RSU) mounted over trucks, can achieve the expected \emph{high data-rate} along with the \emph{dynamic adaptation} to varying demands. Toward designing such \emph{relay-assisted} wireless networks, two \emph{challenges} are considered in this paper. The first challenge focuses on maximizing the \emph{network} flow rate through the optimal \emph{positioning} of relays, which can enable \emph{simultaneous} relay-assisted parallel routes. The second challenge focuses on maximizing the \emph{link} data rate, by designing multi-antenna \emph{beamforming codebooks} that depend on relay positions and spatially-correlated wireless channels. In this paper, we propose \emph{brain-inspired geometric-based} approaches to tackle these two challenges. 

\subsection{Relay Positioning and Maximum Flow}
Maximizing the algebraic connectivity of network graphs have been utilized in finding positions of 2-dimensional (2-d) relays~\cite{Ibrahim_TWC_Connectivity_2009} or 3-d unmanned aerial vehicles (UAVs)~\cite{Mai_2019_Elsevier_UAV_IAB, rahmati2019dynamic}. While maximizing the algebraic connectivity will naturally increase the network flow rate~\cite{rahmati2019dynamic}, it does not achieve the maximum flow rate, as we will show later in this paper. Therefore, we aim to find an alternative optimization metric that can be utilized in positioning relays toward achieving higher network flow rate. To do so, we turn our attention to brain networks and \emph{Riemannian geometry}~\cite{2019_Intro_RiemGeometry}. 

Riemannian geometry has been considered in classifying functional connectivity patterns associated with unique brain tasks (e.g., memory or subtraction)~\cite{2016_Riem_Brain_Decoding}. Such brain classification serves as the main \emph{inspiration} for this paper as follows. Having two functional connectivity patterns that are distinguishable from each other over Riemannian manifolds~\cite{2018_Lee_Book_RiemManifolds} resembles having two \emph{parallel} data flows, which in turn leads to higher network flow rate. We note that covariance matrices of connectivity paths are represented over Riemannian manifolds given their \emph{symmetric positive definite} (SPD) characteristics. Consequently, Riemannian metrics, such as the Log-Euclidean metric (LEM)~\cite{2006_LEM_Arsigny}, have been utilized for task classification.  

In this paper, we geometrically represent regularized \emph{Laplacian} matrices of relay-dependent network graphs, which are SPD ones, over Riemannian manifolds. Consequently and inspired by the LEM-based brain-tasks classification, we identify the optimum relay positions as the ones achieving maximum LEM, compared to baseline network with no relays. We show that the proposed LEM-based relay positioning scheme almost achieves the \emph{maximum flow rate} and can serve as a low-complexity solution for the maximum flow problem~\cite{Edmonds_MFP}. Moreover, we identify parallel (independent) multi-hop routes as the ones with maximum LEM among each other.

\subsection{Beamforming Codebook Design}
As the maximum \emph{network} flow rate is achieved, through the LEM-based relay positioning, we turn our attention to maximizing the relay-user \emph{link} rate. Each optimally-positioned multi-antenna relay will communicate with each of its adjacent users by first estimating its channel vector and then assigning a suitable beamforming codeword. Generally, the relative position of each relay to its users will vary from one relay to the other. Therefore and taking into consideration the practical scenario of \emph{spatially-correlated fading channels}~\cite{2016_Debbah_MassiveMIMO_Covariance}, we aim to design a \emph{unique} beamforming codebook for each relay. 

In designing relay-dependent beamforming codebooks, we represent channel covariance matrices of relay-user spatially-correlated fading channels, which are SPD ones, over Riemannian manifolds. Each user channels follow an exponential correlated fading model~\cite{Clerckx_2008}, which depends on the user's relative location to the relay. We note that Riemannian geometry has been recently considered in designing beamforming vectors~\cite{2019_Fan_Riem_MassiveMIMO, 2017_Chen_Riem_MUI, 2016_Letaief_Riem_mmWave_Precoding}. While these research works present novel geometric perspectives of beamforming design, they have not utilized the \emph{SPD characteristics} of correlated channels. 

In this paper, we propose to employ \emph{a LEM-based geometric support vector machine} (SVM) model to learn the channel covariance matrices of different users over Riemannian manifold. Once distinct groups of these matrices are identified, beamforming codewords are selected to nearly match these channel groups. Any new estimated channel will be classified to one of the groups, based on LEM distance, and assigned a matched beamforming codeword accordingly. We show that the proposed machine learning model requires small number of training samples to approach the link capacity. 

 
\section{System Model}\label{sec_mod}
In this section, we present brief preliminaries on Riemannian geometry then introduce the system model. 


Topological manifolds are spaces that locally resemble the $N$-d real coordinate space $\mathbb{R}^N$, i.e., they can be locally parameterized by $N$ coordinates. Differential manifolds~\cite{2016_DiffGeom_Carmo} are topological ones with smooth changes of coordinates (maps from $\mathbb{R}^n$ to $\mathbb{R}^n$). Tangent space of a differential manifold at some point is a vector space of all vectors that are tangent to the manifold at that point. \emph{Riemannian} geometry is the study of Riemannian manifolds~\cite{2019_Intro_RiemGeometry}, which are differential manifolds with some metric. A Riemannian metric determines an \emph{inner product} on each tangent space, and it measures the \emph{distances} or angles of curves on the Riemannian manifold. Finally, SPD matrices lie on Riemannian manifold and LEM is a valid Riemannian metric for SPD matrices~\cite{2006_LEM_Arsigny}.

A given network can be represented as an undirected finite graph $G(V,E)$, where $V=\{v_1, v_2, \cdots, v_n\}$ is the set of all $n$ nodes and $E$ is the set of all $m$ edges. 
Considering the standard \emph{disk model}, two nodes are connected if their inter-distance is less than a specific threshold $R$. For an edge $l$, $1 \leq l \leq m$, connecting nodes $\{v_i,v_j\}\in V$, define the edge vector ${\mathbf{a_{l}}} \in \mathbb{R}^n$, where the $i$-th and $j$-th are given by $a_{l,i} = 1$ and $a_{l,j} = -1$, respectively, and the rest is zero. The incidence matrix $\mathbf{A} \in {\mathbb{R}^{n \times m}}$ of the graph $G$ is the matrix with $l$-th column given by $\mathbf{a_{l}}$. The \emph{Laplacian} matrix $\mathbf{L} \in \mathbb{R}^{n \times n}$ is defined as $\mathbf{L} = \mathbf{A}\,\mathbf{A}^T$, where $T$ denotes matrix transposition. 
Laplacian matrices are \emph{positive semi-definite}, and their second smallest eigenvalue, $\lambda_2(\mathbf{L})$, is the graph \emph{algebraic connectivity}~\cite{2006_Ghosh_Fiedler}.

Given that the Laplacian matrices are positive semi-definite, a simple regularization step~\cite{2015_Riem_Brain_Class} is implemented to produce a \emph{regularized} SPD Laplacian matrix as 
\begin{equation}
\mathbf{S} = \mathbf{L} + \gamma \, \mathbf{I} =  \mathbf{A}\,\mathbf{A}^T+ \gamma \, \mathbf{I}  \,,
\label{S_matrix}
\end{equation}
where $\mathbf{I}$ is the $n \times n$ identity matrix and $\gamma$ is an arbitrary small scalar (e.g., $\gamma=0.5$). The regularized SPD Laplacian matrix $\mathbf{S}$ lies on Riemannian manifold, and
the LEM between two SPD matrices, $\mathbf{S}_1$ and $\mathbf{S}_2$, can be calculated as~\cite{2006_LEM_Arsigny}
\begin{equation}
\mathcal{D}(\mathbf{S}_1,\mathbf{S}_2)= ||\log(\mathbf{S}_1) - \log(\mathbf{S}_2)||_F^2 \;, 
\label{eqn_LEM}
\end{equation}
where $||\,.\,||_F$ denotes the matrix Frobenius norm.


\section{Relay-based Maximum Flow Problem Formulation} \label{prob_form}
In this section, we formulate the problem of relay positioning as a maximum flow problem. 

We first assume that there exist only $Z$ candidate locations for the deployment of the available $K$ relays, where $K < Z$. Let $p_k$ be the $(x,y)$ position of the $k$-th relay and $\mathbf{P} = [p_1 \, p_2, \cdots, p_K]^T$ be the $K \times 2$ matrix containing positions of all $K$ relays. Deploying a relay in a potential location creates edges between two or more network nodes that are within distance $R$ of the relay location (disk model). Consequently, new edges are added to the original network leading to a new set of edges, denoted as $E(\mathbf{P})$. Furthermore the capacity of link $(i,j) \in E(\mathbf{P})$ between nodes $\{v_i, v_j\} \in V$, denoted as $f_{i,j}$, is either $1$ if their inter-distance is less than the disk radius $R$, or $0$ otherwise. 

For a given source node, $s \in V$, and relay positions, $\mathbf{P}$, the maximum flow problem is formulated as
\begin{align} 
\max  & \; \; f(s, \mathbf{P})\;= \sum_{j:(s,j)\in E(\mathbf{P})} f_{s,j} \;,\nonumber \\
 \text{s.t.} &\; \sum_{i:(i,j)\in E(\mathbf{P})} f_{i,j} - \sum_{u:(j,u)\in E(\mathbf{P})} f_{j,u} = 0 \;, \; \forall j \in V \backslash \{s,d\} \,, \nonumber \\
& f_{i,j} = \{0,1\}\;, \; \forall (i,j) \in E(\mathbf{P}), 
\label{opt_MFP}
\end{align}
in which we aim to maximize the amount of flow generated from the source, $s$, towards its destination, $d \in V$, subject to both conservation and capacity conditions. 
By considering every node in the graph as a potential source node, the average maximum flow rate of the network is computed as $f(\mathbf{P}) = \frac{1}{n} \, \sum_{s \in V} f(s,\mathbf{P})$. The optimum $K \times 2$ positions matrix, $\mathbf{P}^*$, is the one achieving maximum value of $f(\mathbf{P})$, i.e., 
\begin{equation}
\mathbf{P}^* = \argmax_{\mathbf{P}} \frac{1}{n} \; \sum_{s \in V} \sum_{j:(s,j)\in E(\mathbf{P})} f_{s,j} \,.
\label{opt_L}
\end{equation}
Calculating the maximum flow for a given source, $s$, and relay positions, $\mathbf{P}$, requires complexity of $\mathcal{O} (|V|\, |E(\mathbf{P})|^2)$ using the Edmonds–Karp algorithm~\cite{Edmonds_MFP}. In the next section, we show how such high-complex problem can be mapped to a lower-complexity one.

\section{LEM-based Relay Positioning Scheme} \label{solution}
In this section, we introduce our proposed brain-inspired problem transformation to be addressed through Riemannian geometry, then we describe the proposed solution. 

\subsection{Relay Positioning through Brain-inspired Geometric Lens} 
\begin{figure}[htbp]
	\centerline{
		\includegraphics[width=0.4\textwidth]{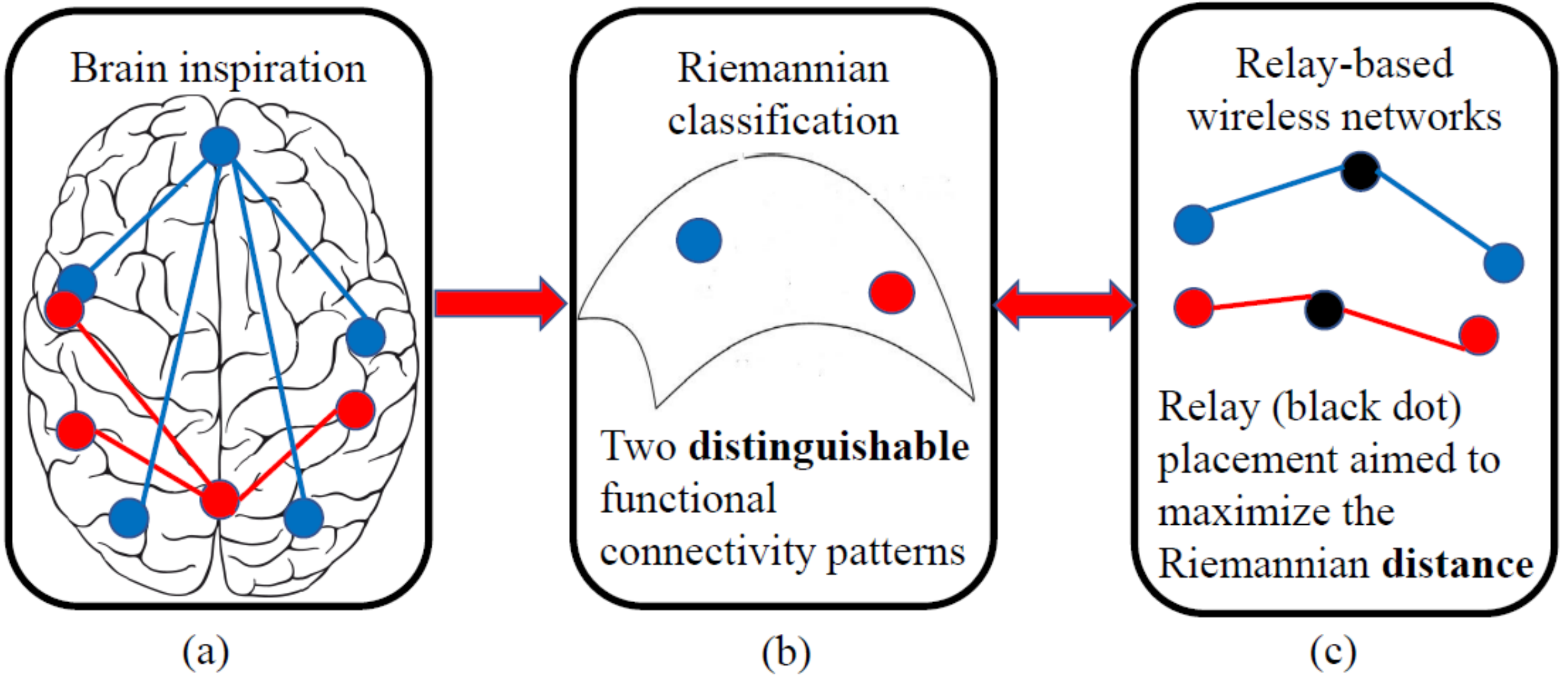}}
	\caption{\small Brain-inspired problem transformation.} 
	\label{fig_brain_inspired}
	\vspace{-0.2in}
\end{figure}

Fig.~\ref{fig_brain_inspired} shows the transformation from a maximum flow problem to a geometric-based one, which stems from functional connectivity analysis in brain networks. Fig.~\ref{fig_brain_inspired}~(a) depicts a simplified version of the results presented in~\cite{2016_Riem_Brain_Decoding}, which indicates that different brain tasks such as memory and subtraction have \emph{distinguishable} data flows among brain regions. On one hand, having independent data paths, which can be seen as a multiple-source multiple-sink maximum flow problem~\cite{1989_Miller_MSMS_Graph}, increases the network flow rate. On the other hand and as shown in Fig.~\ref{fig_brain_inspired}~(b), these two paths are represented as two \emph{separable points} over Riemannian manifolds. Such tasks classification is possible by considering distance-based Riemannian metric such as LEM~(\ref{eqn_LEM}). 

Therefore increasing the network flow rate, via enabling multiple independent paths, can be equivalent to increasing the LEM among the paths' geometric representations over Riemannian manifold. Consequently and as shown in Fig.~\ref{fig_brain_inspired}~(c), the optimum positions of relays can be defined as the ones achieving maximum LEM as compared to the baseline (no-relay) network scenario. In other words, potential relays that result in maximum LEM will lead to independent (i.e., vertex-disjoint or edge-disjoint) paths and hence higher network rate. 


As any potential relay locations matrix, $\mathbf{P}$ results in new edge matrix $E(\mathbf{P})$, then the equivalent regularized Laplacian matrix, $\mathbf{S}_\mathbf{P}$, can be computed as in (\ref{S_matrix}) using the updated edge set $E(\mathbf{P})$. We note that $\mathbf{S}_\mathbf{P}$ is an SPD matrix,  represented as a point on the Riemannian manifold. Therefore, the impact of adding $K$ relays at locations $\mathbf{P}$ can be measured by computing the LEM as $\mathcal{D}(\mathbf{S}_\mathbf{P},\mathbf{S}_b)$, where $\mathbf{S}_b$ represents the regularized Laplacian matrix of the baseline network. The optimum relay locations, $\mathbf{P}^*$, is the one achieving maximum LEM value. In other words, the optimization problem for the relay deployment in (\ref{opt_L}) is transformed to a geometric-based equivalent one as
\begin{equation}
\mathbf{P}^* = \argmax_{\mathbf{P}} \mathcal{D}(\mathbf{S}_\mathbf{P},\mathbf{S}_b) \,.
\label{opt_L_LEM}
\end{equation}

\subsection{LEM-based Relay Positioning and Parallel Routing} \label{relay-to-relay}

The optimum relay positions can be found by solving (\ref{opt_L_LEM}), which can be efficiently computed using optimization techniques over manifolds~\cite{2009_Absil_RiemGeometry} and also geodesically convex optimization~\cite{2016_MIT_Geodesic_Convex} approaches. However in this paper and as a preliminary proof of concept, we have used iterative exhaustive search approach. More specifically, the optimum position of the first relay is determined by choosing the location that maximizes the LEM, compared to the baseline network with no relays. In other words, exhaustive search is conducted over all $Z$ potential locations and the optimum position is the one satisfying (\ref{opt_L_LEM}). Once the first relay is chosen, it is added to the baseline network and the same exhaustive search is repeated to find the best position for the second relay. Such algorithm continues by adding one relay at a time, until all $K$ relays have been optimally positioned.

Once all relays are optimally positioned, clusters are formed around each relay, which acts as a cluster head. Each network node is then associated with a cluster, based on its shortest distance towards the cluster head. Multi-hop communication between any two non-adjacent nodes, $\{v_i, v_j\} \in V$ occurs through relays. Let $W$ denote the set of all combinations of relay-to-relay routes. Furthermore, each possible route, $R_a$, where $a \in W$, can be represented as a point on Riemannian manifold with regularized Laplacian matrix, $\mathbf{S}_a$. The LEM between any two routes, $\{R_a, R_b\}$, is computed as $\mathcal{D}(\mathbf{S}_a,\mathbf{S}_b).$

Towards increasing the network flow rate, we aim to establish multiple parallel paths (routes) among different relays, enabling parallel cluster-to-cluster communication. Parallel routes can be practically defined as the ones with minimum number of overlapping nodes or edges. Consequently and given the brain inspiration, discussed in Fig.~\ref{fig_brain_inspired}, we propose to identify parallel routes as the ones having \emph{maximum LEM} over Riemannian manifold. So data packets can \emph{simultaneously} traverse two optimal relay-to-relay routes $\{R_a^*, R_b^*\}$ given that 
\begin{equation}
\{R_a^*, R_b^*\} = \argmax_{ \{a,b\} \in W} \mathcal{D}(\mathbf{S}_a,\mathbf{S}_b) \;.  
\label{eqn_LEM_routing}
\end{equation}
As a preliminary proof of concept, exhaustive search can be conducted to calculate the LEM among all relay-to-relay routes, and the two routes satisfying (\ref{eqn_LEM_routing}) are chosen. In the future, we will consider alternative approaches in solving (\ref{eqn_LEM_routing}).

\section{Geometric Machine Learning for Beamforming Codebook Design} \label{sol_beamforming}
As Section~\ref{solution} focused on optimal relay positioning and inter-cluster (relay-to-relay) multi-hop communication, this section completes the remaining link by focusing on intra-cluster (relay-to-user) communication. 

\subsection{Spatially-correlated Channel Modeling}
We consider having multiple antennas, $M$, for each relay, while single antenna for each user (network node). It is often assumed that multiple-antenna channels are independent and hence their covariance matrices are scaled version of the identity matrix. However, such assumption is not a practical one, as multiple-antenna channels are generally spatially-correlated~\cite{2016_Debbah_MassiveMIMO_Covariance}. A multiple-input single-output (MISO) channel between a given relay and its user $u$, denoted as $\mathbf{h}_{u}$, can be modeled as a correlated Rayleigh fading channel vector with covariance matrix $\mathbf{Q}_{ku} \in \mathbb{C}^{M \times M}$, i.e.,  $\mathbf{h}_{u}  \sim \mathbb{CN}(\mathbf{0}, \mathbf{Q}_{u})$. Covariance matrices can be generated according to the Clerckx exponential correlation model~\cite{Clerckx_2008}, which depends on the inter-antenna spacing as well as on a \emph{phase} component that is uniformly-distributed over $[0, 2\, \pi]$ to reflect the user's location.

\begin{table}[htbp]
	\vspace{0.1in}	
		\caption{\small Network simulation parameters.}
	\vspace{-0.1in}	
		\begin{center}
		\begin{tabular}{|l|l|}
			\hline
			\textbf{Parameter}&\textbf{Value} \\
			\hline
			Deployment area & $6 \times 6$ \\
			\hline
			Disk model radius ($R$) & $2$ \\
			\hline
			Number of network nodes ($n$) & $20$ \\
			\hline
			Number of potential relay positions ($Z$) & $16$ \\
			\hline
		\end{tabular}
		\label{table_sim_parameters}
	\end{center}
	\vspace{-0.2in}	
\end{table}

\subsection{Geometric Machine Learning}
Generally if the covariance matrices of the spatially-correlated channels are known apriori, beamforming codebook can be designed accordingly. For example, one codeword can be matched to the angular phase of a given covariance matrix. However covariance matrices of relay-user correlated channels are not known apriori, as they depend on the user locations with respect to the optimally-positioned relay~\cite{Clerckx_2008}. Consequently, the beamforming codebook for each relay cannot be designed beforehand, and it needs to be learned based on the user channels within each relay's cluster. Therefore, our goal is to learn the correlation characteristics of each user's channel. 

For simplicity of explanation, we assume two users, $u=\{1,2\}$, and each one follows a different exponential correlation model, as it depends on its unique location. As our goal is to learn the covariance matrix of each user's channel, we turn our attention to learning the $(\mathbf{h}_{u} \, \mathbf{h}_{u}^H)$ matrix for each user, where $H$ denotes matrix hermitian, as opposed to learning the channel vector $\mathbf{h}_{u}$ itself. As the $(\mathbf{h}_{u} \, \mathbf{h}_{u}^H)$ matrix is an $M \times M$ SPD one, it can be represented as a point over Riemannian manifold. Consequently, learning the two covariance matrices can be conducted over Riemannian manifolds, as opposed to conventional Euclidean spaces. In classifying between the two users, the LEM Riemannian metric will be utilized.

In this paper, we propose a \emph{geometric machine learning} approach to learn the covariance matrix of each user by applying the standard SVM model over Riemannian manifold. The proposed geometric SVM classifies the $M \times M$ $\mathbf{h}_{u} \, \mathbf{h}_{u}^H$ SPD channel matrices, for $u=\{1,2\}$, into two groups using the LEM Riemannian distance. Once each of these two groups are constructed, two codewords matching the angles of these two learned covariance matrices are identified. In the testing phase and for any new estimated relay-user channel, it will be classified into one of the two groups and then assigned the corresponding group-specific beamforming codeword. For example, let $\mathbf{h}_{t}$ be a newly estimated channel and it was classified to the $u=1$ group with codeword $\mathbf{c}_1$. The achievable link rate for this channel will be equal to $R_t= \log_2(1+ \text{SNR} \, |\mathbf{h}_{t}^H \, \mathbf{c}_1|^2)$, where SNR is ratio between the signal power to the noise variance. 
 
Unlike the recent works on using deep learning for channel estimation (e.g., \cite{gao2019deep}), in which channel learning happens over Euclidean spaces, our proposed \emph{geometric} machine learning approach is tailored to the practical spatially-correlated multiple-antenna channels by learning such SPD matrices over geometric Riemannian manifold. Finally, we point out that we have utilized basic machine learning schemes, such as SVM, as a proof of concept in this paper. However, advanced geometric deep learning algorithms will be utilized in the future.

\begin{figure}[htbp]
	\vspace{-0.0in}
	\centerline{
		\includegraphics[width=0.42\textwidth]{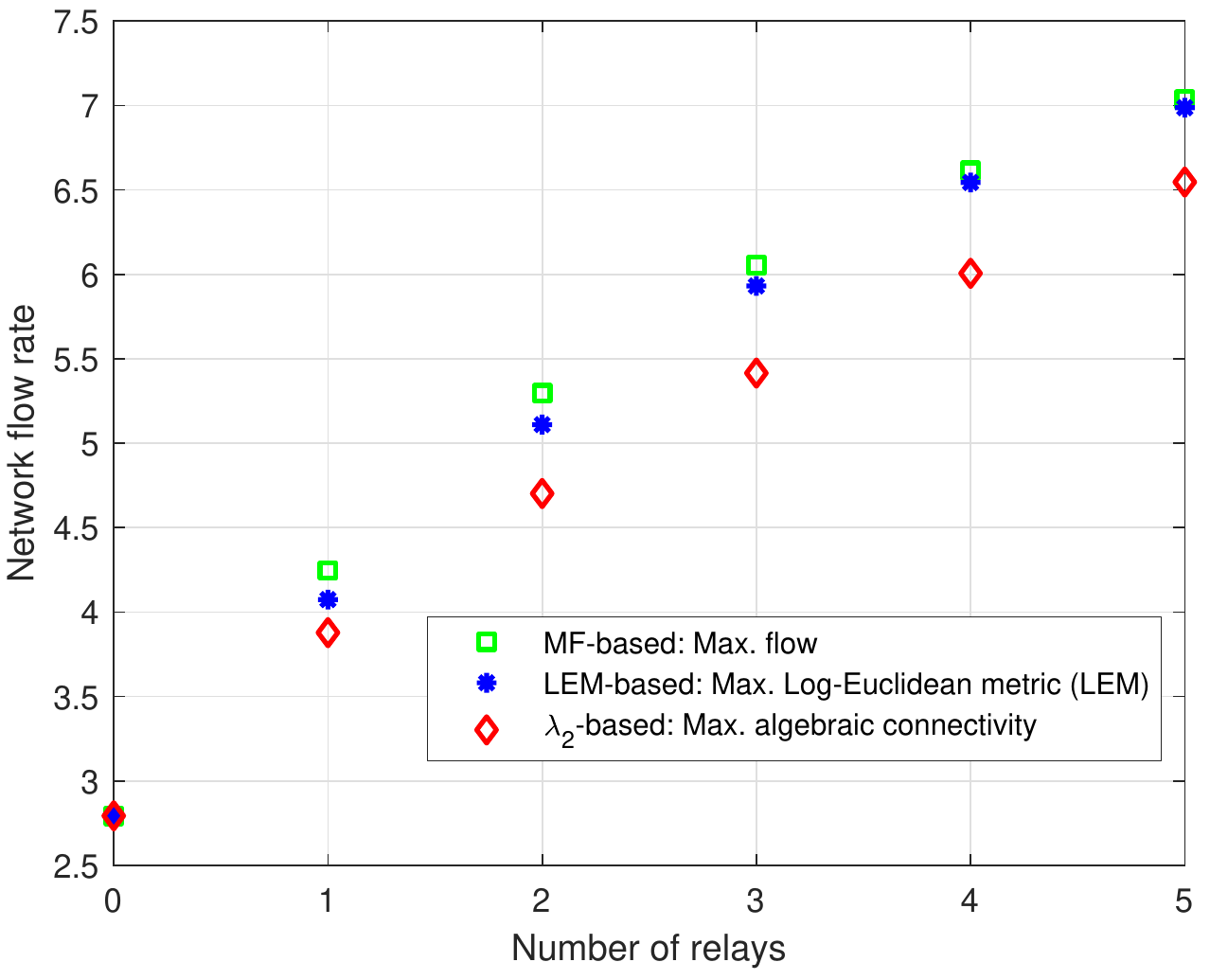}} \vspace{-0.05in}
	\caption{\small Average network flow rate achieved by different relay-positioning optimization metrics.}
	\label{fig_MF_conn}
	\vspace{-0.1in}
\end{figure}

\section{Simulation Results} \label{sim_results}
In this section, we present simulation results of the proposed relay-positioning and geometric channel learning schemes. 

\subsection{LEM-based Relay Positioning and Maximum Flow}
Let \emph{$\lambda_2$-based} scheme be the one finding relay positions by maximizing the algebraic connectivity of the graph (e.g., as in~\cite{rahmati2019dynamic}), while \emph{MF-based} scheme is the one positioning relays that achieve average maximum network flow rate~\cite{Edmonds_MFP}. The relay positions in both cases, along with the proposed LEM-based scheme, were found through exhaustive search of all possible relay locations and finding the location vector maximizing the metric of interest in each case. The main network simulation parameters are included in Table~\ref{table_sim_parameters}. 

Fig.~\ref{fig_MF_conn} shows the achievable average network flow rate by all relay-positioning schemes for $K = 1$ to $5$ relays. As shown, the {$\lambda_2$-based} scheme has a loss of $9\%$ at $K=4$ relays compared to the MF-based one, and this is our motivation to find an alternative optimization metric. Indeed, we find that the proposed LEM-based relay-positioning scheme achieves smaller gap of less than $1\%$ at $K=4$ relays, compared to the \emph{high-complexity} MF-based one. Equally important, our proposed LEM-based scheme has improved the network flow rate of the $\lambda_2$-based one by $9\%$ at $K=4$ relays, while requiring the \emph{same low-complexity} computation. Such gain is simply due to utilizing the regularized Laplacian matrix for calculating the LEM distances over Riemannian manifold, as opposed to computing its second smallest eigenvalue $\lambda_2$.
 
\begin{figure}[htbp]
	\centerline{
		\includegraphics[width=0.43\textwidth]{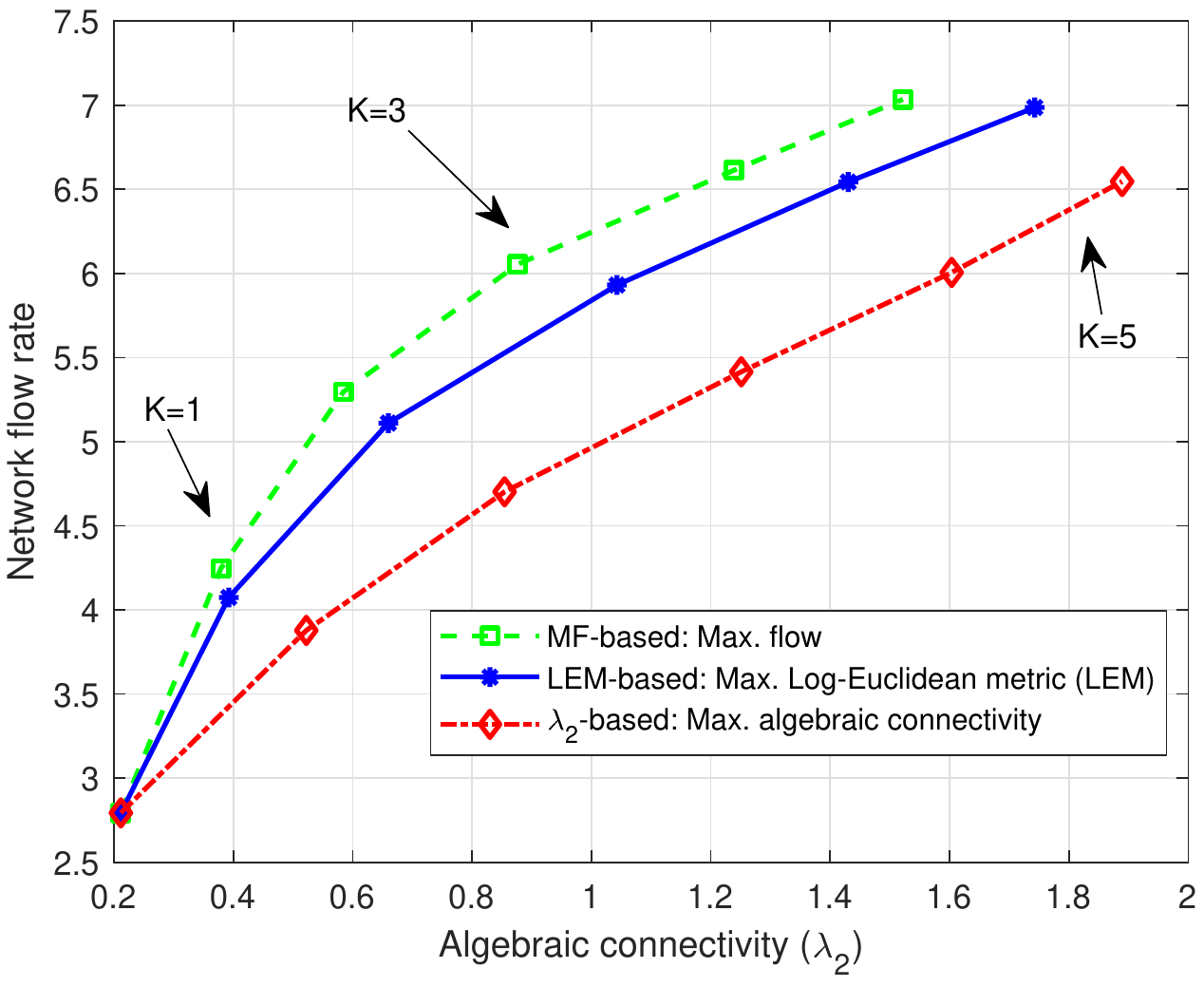}} 
	\caption{\small Average network flow rate versus algebraic connectivity. Markers denote performance at $K=0$ to $5$ relays for different relay-positioning optimization metrics.}
	\label{fig_MF_vs_Lambda}
\end{figure}
While higher network flow rate is of great importance, the \emph{robustness} of networks, measured in terms of its connectivity degree, is also of equal importance. Fig.~\ref{fig_MF_vs_Lambda} depicts the achievable network flow and algebraic connectivity for the three relay-positioning metrics. For given number of relays ($1 \leq K \leq 5$) and as expected, the maximum flow is achieved by the MF-based positioning algorithm, while the maximum algebraic connectivity is achieved by the $\lambda_2$-based one. Interestingly, Fig.~\ref{fig_MF_vs_Lambda} shows that the performance of proposed LEM-based scheme lies in between these two benchmark schemes. 

In other words, the proposed LEM-based scheme achieves a unique and balanced \emph{tradoeff} performance between the network flow rate and algebraic connectivity, which is not achievable by any other scheme. Such unique performance is due to the novel consideration of brain-inspired Riemannian geometry in addressing the relay positioning problem. We point out that the line segment between any two markers in~Fig.~\ref{fig_MF_vs_Lambda} is achieved by standard time sharing strategy across two different numbers of deployed relays.


\begin{table}[!t]
	\captionof{table}{\small Average number of overlapping nodes and edges among parallel routes, for different number of $K$ relays.}\label{table_cong}
	\def\arraystretch{2}
	\setlength{\tabcolsep}{7.5pt}  
	\centering
	\begin{tabular}{|c|c|c|c|c|}
		\cline{2-5}
		\multicolumn{1}{c|}{} &\multicolumn{2}{|c|}{Overlapping Nodes} & \multicolumn{2}{|c|}{Overlapping Edges}\\ 
		\cline{2-5} \hline
		\multicolumn{1}{|c|}{$K$} & LEM & MF & LEM & MF \\ 
		\cline{2-5}\hline
		3 & 1.01 & 1.31 & 0.01 & 0.33 \\ \hline
		4 & 0.35 & 0.31 &  0.04 & 0.06   \\ \hline
		5 & 0.25 & 0.31 &   0 & 0.03  \\ \hline
	\end{tabular}
	\vspace{-0.15in}
\end{table}

Upon optimally-positioning relays, we consider multi-hop relay-to-relay communication. As our goal is to have independent routes that can occur simultaneously, Table~\ref{table_cong} presents the average number of overlapping nodes and edges among inter-cluster (inter-relay) routes for $K=3$ to $5$ relays. We note that the minimum value of $K=3$ is chosen to allow two parallel routes among all relays Otherwise, there will be only one route among relay(s). Furthermore, routing path between each two relays was determined using the standard Dijkstra's shortest path algorithm~\cite{dijkstra1959note}. As proposed in (\ref{eqn_LEM_routing}), the two chosen parallel routes are the ones with maximum LEM among all potential relay-to-relay routing paths. 
Table~\ref{table_cong} shows that the LEM-based routing scheme achieves comparable congestion results to that achieved by the MF-based one. In other words, the proposed LEM-based relay-positioning and routing schemes enable parallel routing with minimal congestion levels at nodes and edges, similar to the high-complexity MF-based one.

\begin{figure}[htbp]
	\centerline{
		\includegraphics[width=0.43\textwidth]{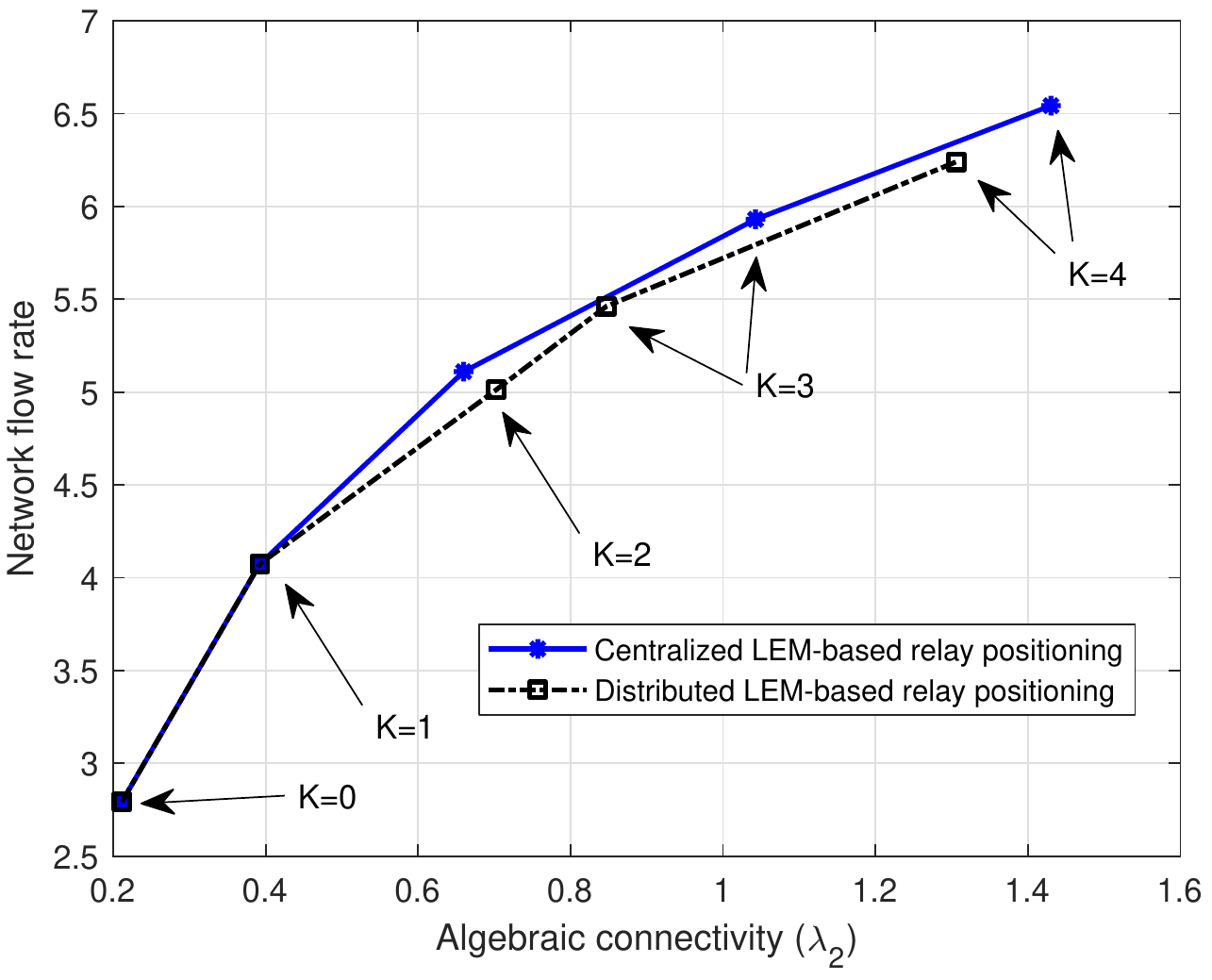}} 
	\caption{\small Maximum flow versus algebraic connectivity for distributed LEM implementation considering $K=0$ to $5$ relays.}
	\label{fig_distributed_LEM}
	\vspace{-0.1in}
\end{figure}
The previous results were done assuming a centralized control unit that is aware of all network node locations and it deploys one relay at a time through exhaustive search. An alternative localized approach is the \emph{distributed} one, which partitions the area of interest into a number of non-overlapping regions. Each region has its own local control unit that is aware of node locations within its smaller region. Furthermore, each local unit positions one relay within its region. For example if $K=4$ relays, the area of interest is divided into $4$ equal quarters, and one relay is deployed in each quarter following the LEM-based scheme, presented in Section~\ref{solution}. Fig.~\ref{fig_distributed_LEM} depicts the distributed implementation of LEM-based relay positioning and its performance with respect to the centralized one, presented earlier in Fig.~\ref{fig_MF_vs_Lambda}. As shown, the performance loss due to distributed implementation at $K=4$ is $4.6\%$ in network flow rate and $8.7\%$ in algebraic connectivity.

\subsection{Beamforming Codebook Design for Correlated Fading}
We assume two users of spatially-correlated channels, each one follows the Clerckx exponential correlation model~\cite{Clerckx_2008} with unique phase values of $\pi$ and $0$. Assuming $M \in \{2,4\}$ antennas, then the correlation covariance matrix of user $u \in \{1,2\}$, denoted as $Q_u^M$, can be written as
\begin{equation}
Q_u^2 = \begin{bmatrix} 
1 & t_u  \\
t_u^* & 1 \\			
\end{bmatrix} \; \; \;,
Q_u^4 = \begin{bmatrix} 
1 & t_u & t_u^2 &  t_u^3 \\
t_u^* & 1 & t_u & t_u^2 \\
{t_u^*}^2 &	t_u^* & 1 & t_u  \\
{t_u^*}^3 & {t_u^*}^2 &	t_u^* & 1 \\			
\end{bmatrix} \;,
\label{corr_matrix}
\end{equation}
where $t_u$ is the transmit correlation coefficient for user $u$. We assume the two users have the same absolute value~\cite{Clerckx_2008}, for example, $|t_1|=|t_2|=0.5$. On the contrary, the phases of the transmit correlation coefficients are different as $\angle t_1 = \pi$ and $\angle t_2 = 0$. Uniform planner array (UPA) deployment of $M \times 1$ antenna structure is deployed at the relay. 


The geometric LEM-based SVM was applied over Riemannian manifold using the \emph{``geomstats''}  python package along with its \emph{brain connectome} classification package~\cite{2018_geomstats}. First, we generate total number of $S$ \emph{training} channel samples, which are equally generated from $Q_1^M$ and $Q_2^M$. 
Second, the geometric SVM learns the channel covariance matrices of each user, utilizing the LEM. Geometric SVM learning results in two distinguishable groups of channels. Third, we construct $M \times 2$ codebook, having one codeword matched to covariance matrix of each differentiated group. This is done by choosing directional cosine angles of the UPA that result in maximum capacity for each group of channels. For example given the channel covariance matrices, defined in (\ref{corr_matrix}), the chosen codebooks will be $1/\sqrt{2}\begin{bmatrix}  1 & 1  \\ -1 & 1\\ \end{bmatrix}$ for $M=2$ and $1/2\begin{bmatrix}   1 & -1 & 1 & -1  \\ 1 & 1 & 1 & 1 \\	\end{bmatrix}^T$ for $M=4$ antennas. 

\begin{figure}[htbp]
	\centerline{
		\includegraphics[width=0.42\textwidth]{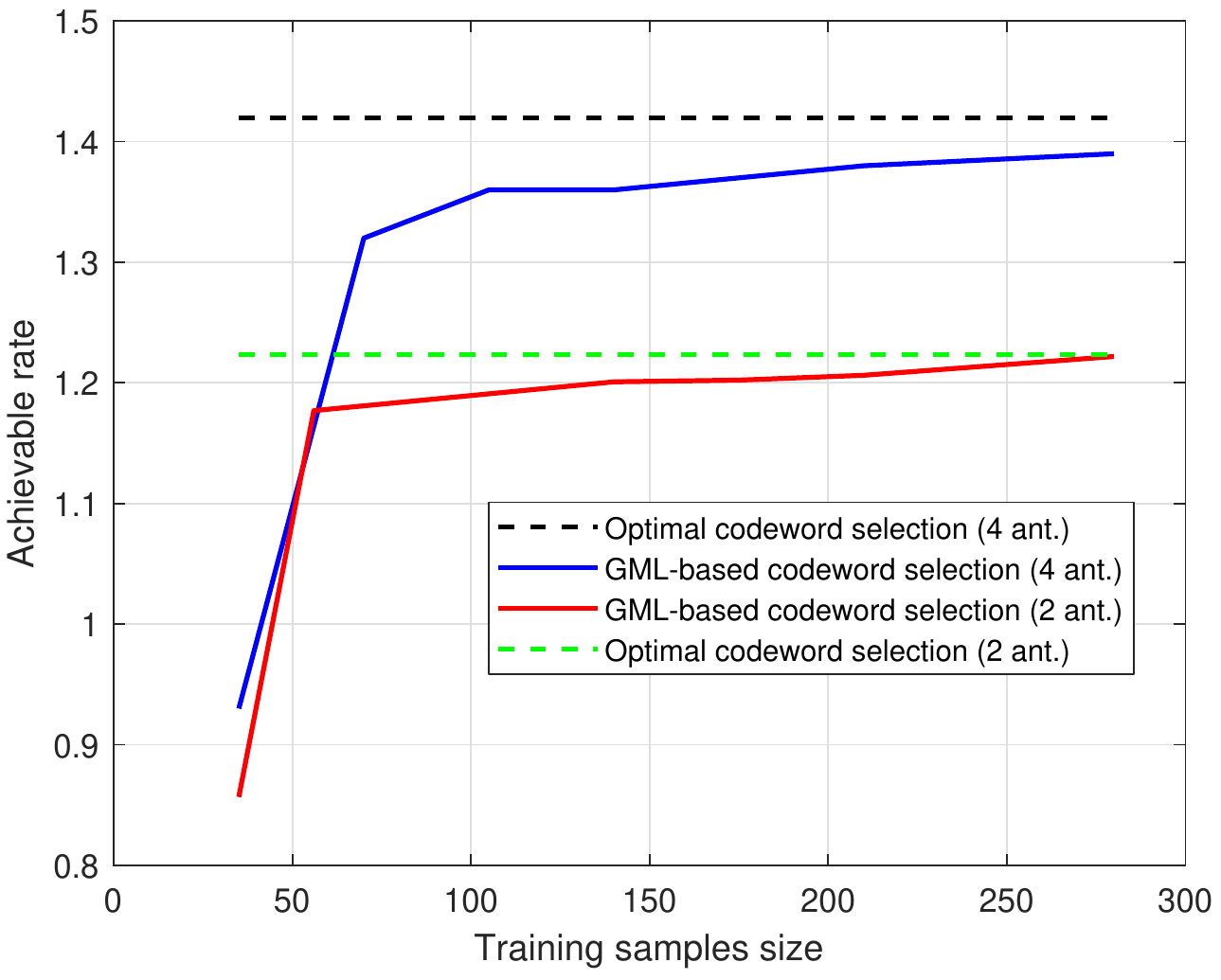}} 	
	\caption{\small Achievable data rate by the geometric machine learning (GML) scheme versus the optimal one for $2$ and $4$ antennas.}
	\label{fig_ML_4ant}
	\vspace{-0.2in}
\end{figure}

\vspace{0.1in}
Fig.~\ref{fig_ML_4ant} depicts the achievable rate of the proposed geometric SVM algorithm for $M=2$ and $4$ antennas, as a function of the training size, $S$. For each $S$ training samples, we generate a \emph{unique} set of $0.4 \, S$ channel samples for testing. We emphasis that the shown rate values are calculated solely based on the testing samples, which have not been used at all in the training phase. It is shown that as the training data size increases, the achievable rate approaches the genie-aided maximum capacity, which is calculated by assigning the best codework to each channel. It is shown in Fig.~\ref{fig_ML_4ant} that more than $90\%$ of the maximum capacity can be achieved by having training size of $S=100$. Finally, Fig.~\ref{fig_ML_4ant} shows that having $M=4$ antennas achieves higher capacity than $M=2$ antennas as more antennas results in additional power gain. 

\section{Conclusion} \label{conc}
In this paper, we have introduced new perspective on designing wireless networks through geometric lens. First, we have utilized Log-Euclidean metric (LEM) for relay positioning over Riemannian manifolds. The proposed LEM-based scheme approaches the maximum flow rate, and it also achieves a unique tradeoff between maximum flow rate and robustness (algebraic connectivity). Second, we have shown that LEM-based inter-relay parallel routes occur with minimal overlapping of nodes or edges. Third, we have shown that a distributed implementation of LEM-based relay positioning only losses $4.6\%$ of the network rate, compared to the centralized one. Finally, we have proposed a geometric support vector machine learning model to classify users spatially-correlated fading channels, and choose a beamforming codeword accordingly. We have shown that more than $90\%$ of the optimal capacity can be achieved by having training size of $100$ channels.

\vspace{0.0in}
\footnotesize \setlength{\baselineskip}{15pt}
\bibliographystyle{IEEEbib}
\bibliography{bib/Geometric_GC_bib,bib/Ibrahim}

\begin{thebibliography}{10}

\bibitem{Kibilda_spatio-temporal_traffic}
J.~{Kibilda} and G.~{de Veciana},
\newblock ``{Dynamic Network Densification: Overcoming Spatio-Temporal
  Variability in Wireless Traffic},''
\newblock in {\em 2018 IEEE Global Communications Conference (GLOBECOM)}, 2018.

\bibitem{Ibrahim_TWC_Connectivity_2009}
A.S. Ibrahim, K.G. Seddik, and K.J.R. Liu,
\newblock ``{Connectivity-Aware Network Maintenance and Repair via Relays
  Deployment},''
\newblock {\em IEEE Trans. on Wireless Communication}, vol. 8, pp. 356--366,
  Jan. 2009.

\bibitem{Mai_2019_Elsevier_UAV_IAB}
M.~A. Abdel-Malek, A.~S. Ibrahim, M.~Mokhtar, and K.~Akkaya,
\newblock ``{UAV Positioning for Out-of-Band Integrated Access and Backhaul
  Millimeter Wave Network},''
\newblock {\em Physical Communication}, vol. 35, 2019.

\bibitem{rahmati2019dynamic}
A.~Rahmati, X.~He, I.~Guvenc, and H.~Dai,
\newblock ``{Dynamic Mobility-Aware Interference Avoidance for Aerial Base
  Stations in Cognitive Radio Networks},'' 2019.

\bibitem{2019_Intro_RiemGeometry}
S.~Gudmundsson,
\newblock {\em {An Introduction to Riemannian Geometry}},
\newblock Lund University, 2019.

\bibitem{2016_Riem_Brain_Decoding}
B.~{Ng}, G.~{Varoquaux}, J.~B. {Poline}, M.~{Greicius}, and B.~{Thirion},
\newblock ``{Transport on Riemannian Manifold for Connectivity-Based Brain
  Decoding},''
\newblock {\em IEEE Trans. on Med. Imaging}, vol. 35, no. 1, pp. 208--216, Jan
  2016.

\bibitem{2018_Lee_Book_RiemManifolds}
J.~M. Lee,
\newblock {\em {Introduction to Riemannian Manifolds}},
\newblock Springer, 2018.

\bibitem{2006_LEM_Arsigny}
V.~Arsigny, P.~Fillard, X.~Pennec, and N.~Ayache,
\newblock ``{Log-Euclidean Metrics for Fast and Simple Calculus on Diffusion
  Tensors},''
\newblock {\em Magnetic Resonance in Medicine}, vol. 56, no. 2, pp. 411--421,
  2006.

\bibitem{Edmonds_MFP}
J.~Edmonds and R.~M. Karp,
\newblock ``{Theoretical Improvements in Algorithmic Efficiency for Network
  Flow Problems},''
\newblock {\em J. ACM}, vol. 19, no. 2, pp. 248–264, Apr. 1972.

\bibitem{2016_Debbah_MassiveMIMO_Covariance}
E.~{Björnson}, L.~{Sanguinetti}, and M.~{Debbah},
\newblock ``{Massive MIMO with Imperfect Channel Covariance Information},''
\newblock in {\em 2016 50th Asilomar Conf. on Signals, Systems and Computers},
  Nov 2016, pp. 974--978.

\bibitem{Clerckx_2008}
B.~{Clerckx}, G.~{Kim}, and S.~{Kim},
\newblock ``{Correlated Fading in Broadcast MIMO Channels: Curse or
  Blessing?},''
\newblock in {\em IEEE GLOBECOM}, 2008.

\bibitem{2019_Fan_Riem_MassiveMIMO}
W.~{Fan}, C.~{Zhang}, and Y.~{Huang},
\newblock ``{Flat Beam Design for Massive MIMO Systems via Riemannian
  Optimization},''
\newblock {\em IEEE Wireless Communications Letters}, vol. 8, no. 1, pp.
  301--304, Feb 2019.

\bibitem{2017_Chen_Riem_MUI}
J.~{Chen},
\newblock ``{Low-PAPR Precoding Design for Massive Multiuser MIMO Systems via
  Riemannian Manifold Optimization},''
\newblock {\em IEEE Communications Letters}, vol. 21, no. 4, pp. 945--948,
  April 2017.

\bibitem{2016_Letaief_Riem_mmWave_Precoding}
X.~{Yu}, J.~{Shen}, J.~{Zhang}, and K.~B. {Letaief},
\newblock ``{Alternating Minimization Algorithms for Hybrid Precoding in
  Millimeter Wave MIMO Systems},''
\newblock {\em IEEE Journal of Selected Topics in Sig. Proc.}, vol. 10, no. 3,
  2016.

\bibitem{2016_DiffGeom_Carmo}
M.P. do~Carmo,
\newblock {\em {Differential Geometry of Curves and Surfaces: Revised and
  Updated Second Edition}},
\newblock Dover Books on Mathematics. Dover Publications, 2016.

\bibitem{2006_Ghosh_Fiedler}
A.~{Ghosh} and S.~{Boyd},
\newblock ``{Growing Well-connected Graphs},''
\newblock in {\em Proceedings of the 45th IEEE Conference on Decision and
  Control}, Dec 2006, pp. 6605--6611.

\bibitem{2015_Riem_Brain_Class}
L.~{Dodero}, H.~Q. {Minh}, M.~S. {Biagio}, V.~{Murino}, and D.~{Sona},
\newblock ``{Kernel-based Classification for Brain Connectivity Graphs on the
  Riemannian Manifold of Positive Definite Matrices},''
\newblock in {\em 2015 IEEE 12th International Symposium on Biomedical Imaging
  (ISBI)}, April 2015, pp. 42--45.

\bibitem{1989_Miller_MSMS_Graph}
G.~L. {Miller} and J.~{Naor},
\newblock ``{Flow in Planar Graphs with Multiple Sources and Sinks},''
\newblock in {\em 30th Annual Sym. on Foundations of Comp. Science}, 1989.

\bibitem{2009_Absil_RiemGeometry}
P-A Absil, R.~Mahony, and R.~Sepulchre,
\newblock {\em {Optimization Algorithms on Matrix Manifolds}},
\newblock Princeton University Press, 2009.

\bibitem{2016_MIT_Geodesic_Convex}
H.~Zhang and S.~Sra,
\newblock ``{First-order Methods for Geodesically Convex Optimization},''
\newblock in {\em Proceedings of Conference on Learning Theory}, 2016.

\bibitem{gao2019deep}
S.~Gao, P.~Dong, Z.~Pan, and G.~Y. Li,
\newblock ``{Deep Learning based Channel Estimation for Massive MIMO with
  Mixed-Resolution ADCs},'' 2019.

\bibitem{dijkstra1959note}
E.~W. Dijkstra,
\newblock ``{A Note on Two Problems in Connexion with Graphs},''
\newblock {\em Numerische Mathematik}, vol. 1, no. 1, pp. 269--271, 1959.

\bibitem{2018_geomstats}
N.~Miolane, J.~Mathe, C.~Donnat, M.~Jorda, and X.~Pennec,
\newblock ``{geomstats: a Python Package for Riemannian Geometry in Machine
  Learning},'' 2018.

\end{thebibliography}

\end{document}